# Sub-femtosecond synchronization of microwave oscillators with mode-locked Er-fiber lasers

Kwangyun Jung and Jungwon Kim*

*KAIST Institute for Optical Science and Technology and School of Mechanical, Aerospace and Systems Engineering,
Korea Advanced Institute of Science and Technology (KAIST), Daejeon 305-701, South Korea*
*\*Corresponding author: jungwon.kim@kaist.ac.kr*

We synchronize a 8.06-GHz microwave signal from a voltage-controlled oscillator with an optical pulse train from a 77.5-MHz mode-locked Er-fiber laser using a fiber-based optical-microwave phase detector. The residual phase noise between the optical pulse train and the synchronized microwave signal is -133 dBc/Hz (-154 dBc/Hz) at 1 Hz (5 kHz) offset frequency, which results in 838 as integrated rms timing jitter [1 Hz – 1 MHz]. The long-term residual phase drift is 847 as (rms) measured over 2 hours, which reaches $4\times10^{-19}$ fractional frequency instability at 1800 s averaging time. This method has a potential to provide both sub-fs-level short-term phase noise and long-term phase stability in microwave extraction from mode-locked fiber lasers. © 2012 Optical Society of America

Ultralow phase noise microwave sources are important for many scientific and engineering applications such as time-frequency metrology and standards, synchronization of large-scale scientific facilities (such as free-electron lasers), radars, networks, communication, and signal measurement instruments. Various types of ultralow noise microwave sources have been developed for the last 30 years or so, which culminated in state-of-the-art commercial microwave sources such as sapphire-loaded cavity oscillators (SLCOs) [1] and optoelectronic oscillators (OEOs) [2]. Recently, photonic generation of microwave signals from stabilized femtosecond mode-locked lasers has been an active field of research due to the ultralow noise properties of cavity-stabilized cw lasers and mode-locked lasers [3-8]. It has been even shown that solid-state or fiber mode-locked lasers themselves can generate sub-100-as level timing jitter optical pulse trains [9-11].

Although the optical pulse trains can carry ultralow timing jitter below a femtosecond, the transfer of timing stability from the optical domain to the electronic domain is highly nontrivial. When direct photodetection is used, excess phase noise is added in the optical-to-electronic (O/E) conversion process due to nonlinearity, saturation, temperature drift, and amplitude-to-phase conversion in photodiodes [12-14]. In order to address this issue, several approaches have been recently demonstrated. The repetition rate scaling by external interferometers has greatly improved the short-term residual phase noise well below -160 dBc/Hz level [15,16]. On the other hand, to address the long-term phase stability, optical feedback control has enabled sub-fs relative timing drift [5,17].

Although the short-term phase noise or the long-term phase stability has been greatly improved by various methods [5, 8, 15-17], it would be desirable to have a method that can scale down the residual short-term jitter and long-term drift simultaneously. For this task, phase detection method based on phase error-dependent intensity imbalance between the two outputs from the Sagnac loop interferometer [18] can be used. However, the practical applicability and achievable performance were limited by alignment sensitivity, resulting in only 60 fs precision synchronization so far [18]. In this Letter we extend this idea to a robust fiber-based optical-microwave phase detector that enables both ultralow residual short-term (<1 s) phase noise and long-term (>1 h) phase drift below a femtosecond level. An 8.06-GHz microwave signal from a voltage-controlled oscillator (VCO) is synchronized with a 77.5-MHz femtosecond mode-locked Er-fiber laser with out-of-loop residual phase noise of -133 dBc/Hz (-154 dBc/Hz) at 1 Hz (5 kHz) offset frequency, which results in 838 as integrated rms timing jitter [1 Hz – 1 MHz]. The long-term residual out-of-loop rms phase drift is measured to be 847 as (rms) over 2 hours.

For long-term stable, high-precision synchronization between microwave VCOs and mode-locked Er-fiber lasers, we detect and compensate for the phase error between them in the optical domain using a fiber-based optical-microwave phase detector. Figure 1 shows the schematic of the fiber-based optical-microwave phase detector. It is based on the phase detection using the phase error-dependent intensity imbalance between the two outputs from the Sagnac-loop interferometer ($P_1 = P_{in}\cos^2(\Delta\Phi/2)$ and $P_2 = P_{in}\sin^2(\Delta\Phi/2)$, where $P_{in}$ is Sagnac-loop input optical power and $\Delta\Phi$ is the phase difference between counter-propagating pulses in the loop). The polarization-maintaining (PM) fiber loop is built with a 50:50 coupler and a unidirectional high-speed LiNbO$_3$ phase modulator. When a microwave signal with a frequency of integer-multiple to the laser repetition rate is applied to the phase modulator, the phase of optical pulses is modulated according to the temporal position between the optical pulses and the driving microwave signals. The intensity balancing between the two outputs from the Sagnac-loop is obtained when the phase difference $\Delta\Phi$ is set to $\pi/2$. This necessitates a nonreciprocal quarter-wave bias between counter-propagating pulses in the fiber loop in order to lock the optical pulse train at the zero-crossings of the microwave signal. For this, we employ a nonreciprocal quarter-wave bias unit with two Faraday rotators and a quarter-wave plate in the fiber loop [19]. As we use a unidirectional phase modulator, the power difference between the two Sagnac-loop outputs is proportional to the phase error between the optical pulse train and the microwave signal ($\theta_e$ in Fig. 1). When detected by a balanced photodetector, the output voltage signal can be used for precise optical-microwave phase detection. The phase error detection sensitivity is expressed as $K_d = GRP_{avg}\phi_o$ [V/rad], where $G$ is transimpedance gain of the balanced photodetector; $R$ is the responsivity of photodiodes; $P_{avg}$ is the average optical power at the two Sagnac-loop output ports; $\phi_o$ is the modulation depth of the phase modulator.

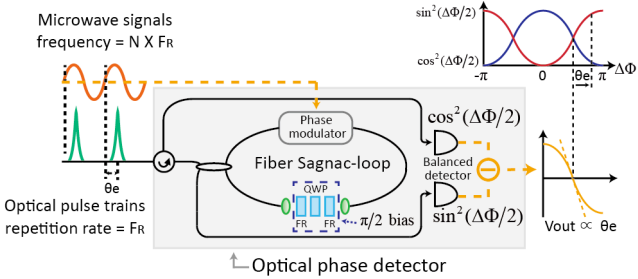

Fig. 1. Fiber-based optical-microwave phase detector.

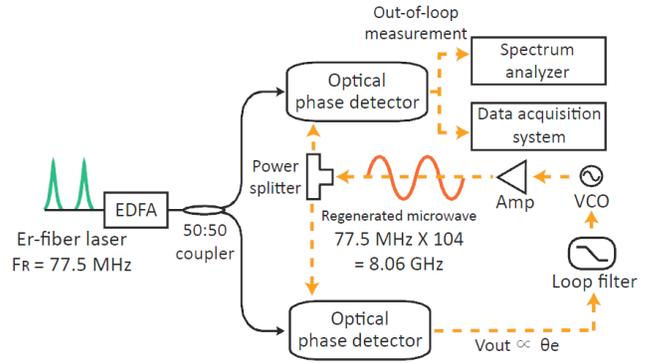

Fig. 2. Optical-microwave synchronization and out-of-loop phase noise/drift measurement set-up.

Figure 2 shows the schematic of the optical-microwave synchronization and the out-of-loop phase measurement setup. Using the fiber-loop-based optical phase detector and a phase-locked loop (PLL) scheme, an 8.06-GHz (the 104th harmonic of the pulse repetition rate) microwave signal from a VCO (Hittite HMC-C200) is synchronized with a 77.5-MHz optical pulse train generated from a home-built, low-jitter mode-locked Er-fiber laser [9]. For the out-of-loop residual phase noise and drift measurement, we built a second optical phase detector. After EDFA and 50:50 coupler, ~40 mW optical power is applied to each optical phase detector. When the VCO is locked to the optical pulse train, ~3.5 mW optical power is detected at each photodiode in the balanced photodetector due to insertion losses of the phase modulator and the bias unit in the fiber loop. Microwave power of +14 dBm from the VCO is amplified to +21 dBm, and +17 dBm is applied to each phase modulator. When these optical and microwave power levels are used, the amplitude-to-phase coefficient of the optical phase detector (at 8.06 GHz carrier) is measured to be 0.06 rad/($\Delta P/P_0$) and 0.3 rad/($\Delta P/P_0$) at 5 kHz and 4 Hz offset frequency, respectively. To attain enough phase margin for the PLL, a combination of lead compensator and proportional-integral (PI) controller is employed. The output of the out-of-loop optical phase detector is applied to the RF spectrum analyzer and the data acquisition board for measuring residual phase noise and drift, respectively.

Figure 3(a) shows the single-sideband (SSB) phase noise measurement results at 8.06 GHz carrier frequency: curve (i) shows the phase noise of the free-running VCO and curve (ii) shows the out-of-loop residual phase noise of the locked VCO. Curve (ii) is subtracted by 3 dB within the locking bandwidth (< ~100 kHz) from the measured data in order to represent the noise of a single system because the measured data is limited by the same additive phase noise in phase detectors within the locking bandwidth. The residual phase noise is -133 dBc/Hz and -154 dBc/Hz level at 1 Hz and 5 kHz offset frequency, respectively, which results in 838 as rms timing jitter integrated from 1 Hz to 1 MHz. Note that the integrated jitter is concentrated in the 100 kHz – 1 MHz range (819 as), whereas the jitter in 1 Hz – 100 kHz contributes only 179 as. The residual phase noise is ultimately limited by the phase noise of the free-running VCO outside the locking bandwidth, thus, it is important to use a high-quality VCO to achieve sub-fs-level synchronization. Curve (iii) is the background noise of the optical phase detector, when only optical pulse train is applied to the phase detector without microwave signal. We believe that the background noise is mainly caused by the amplitude-to-phase conversion in the optical phase detector by imperfect coupling ratio (±1 % error) of the 50:50 coupler in the Sagnac-loop. Curve (iv) shows the in-loop phase error measured in the optical PLL. The measured out-of-loop phase noise is much higher than the projected in-loop phase noise because the PLL performance is limited by the additive phase noise in the optical phase detector, which cannot be suppressed by the loop gain.

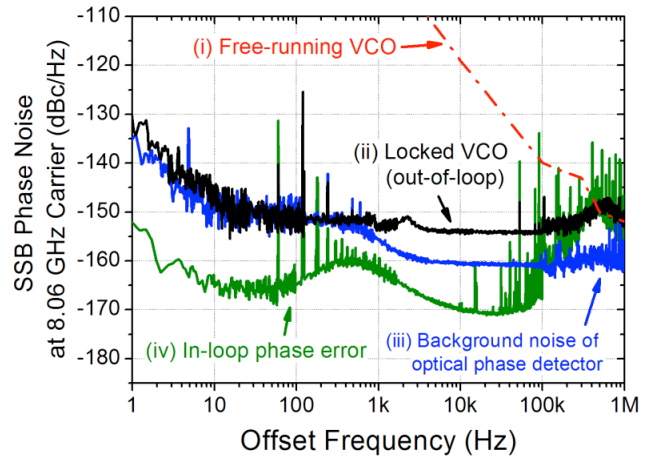

Fig. 3. (a) Signle-sideband (SSB) phase noise measurement results: (i) Phase noise of free-running VCO; (ii) Out-of-loop residual phase noise of VCO locked to the Er-fiber laser (integrated rms timing jitter = 838 as [1 Hz – 1 MHz]); (iii) Background noise of optical phase detector when microwave signal is not applied; (iv) In-loop residual phase noise.

Figure 4 shows the typical out-of-loop long-term phase drift measurement result between the optical pulse trains and the synchronized microwave signals sampled at 2 samples/s with a 100 Hz low-pass filter. For the 2-hour measurement (Fig. 4(a)), the integrated phase drift is 847 as (rms). The slow phase drift is mainly caused by the phase drift in the bias unit (originated from the temperature-dependent birefringence change in the quarter-wave plate) with a temperature coefficient of ~20 fs/K. Based on the measured relative timing drift, we also calculated the fractional frequency instability (in terms of overlapping Allan deviation). Starting from $1.5\times10^{-16}$ at 1 s averaging time, it reaches $4.0\times10^{-19}$ level at 1800 s averaging time (Fig. 4(b)) for a single system.

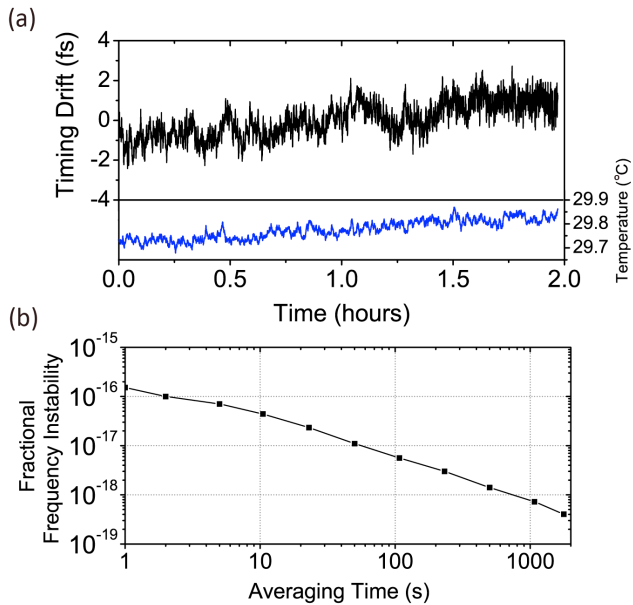

Fig. 4. Long-term residual phase drift measurement result between the optical pulse train and the synchronized microwave signal. (a) Measurement over 2 hours (847 as (rms) drift). (b) Fractional frequency instability (overlapping Allan deviation) based on the measured result in (a).

In summary, we have synchronized an 8.06-GHz microwave oscillator with a 77.5-MHz mode-locked Er-fiber laser with residual phase noise of -133 dBc/Hz (-154 dBc/Hz) at 1 Hz (5 kHz) offset frequency, which results in 838 as integrated rms jitter [1 Hz – 1 MHz]. Long-term phase drift is 847 as (rms) integrated for 2 hours. One of the advantages of this scheme is that there is no restriction on usable repetition rate of mode-locked lasers. Another advantage is that one can obtain stable high-power microwaves from mode-locked lasers without microwave amplification processes by using a high-power, high-quality VCO. One can further build highly-synchronized, ultralow-noise and remotely distributed microwave signal generators for driving accelerating cavities in free-electron lasers by combining the demonstrated PLL system with timing-stabilized fiber links [20]. The fiber-based optical-microwave phase detector itself can be also used for precise measurement of the phase noise of microwave sources or extracting ultrastable microwave signals from optical atomic clocks. As a future work we will measure the absolute phase noise of microwave signals using two independent ultralow jitter mode-locked Er-fiber lasers for the demonstration of ultralow phase noise microwave generators.

We thank Giuseppe Marra for fruitful discussions. This research was supported in part by the Technology Innovation Program funded by the Ministry of Knowledge and Economy (MKE) and the National Research Foundation (NRF) of Korea (2012R1A2A2A01005544).